\documentclass{sigchi-ext}
\usepackage[T1]{fontenc}
\usepackage{textcomp}
\usepackage[scaled=.92]{helvet} 
\usepackage{graphicx} 
\usepackage{balance}  
\usepackage{booktabs} 
\usepackage{ccicons}  
\usepackage{ragged2e} 

\def\plaintitle{SIGCHI Extended Abstracts Sample File: Note Initial
  Caps} 
\def\emptyauthor{}
\def\plainkeywords{Personal health informatics; food journal; eating behaivors; reflection.}

\title{Eat4Thought: A Design of Food Journaling}

\numberofauthors{2}
\author{%
  \alignauthor{%
    \textbf{Yixuan Zhang}\\
    \affaddr{Northeastern University} \\
    \affaddr{Boston, MA 02115, USA} \\
    \email{zhang.yixua@northeastern.edu} } 
    \alignauthor{%
    \textbf{Andrea G. Parker}\\
    \affaddr{Georgia Institute of Technology}\\
    \affaddr{Atlanta, GA 30308, USA}\\ 
    \email{andrea@cc.gatech.edu} }
}

\definecolor{linkColor}{RGB}{6,125,233}
\hypersetup{%
  pdftitle={\plaintitle},
  pdfauthor={\emptyauthor},
  pdfkeywords={\plainkeywords},
  bookmarksnumbered,
  pdfstartview={FitH},
  colorlinks,
  citecolor=black,
  filecolor=black,
  linkcolor=black,
  urlcolor=linkColor,
  breaklinks=true,
}

\begin{document}

\CopyrightYear{2020}
\setcopyright{rightsretained}
    \conferenceinfo{CHI'20 Extended Abstracts,}{April  25--30, 2020, Honolulu, HI, USA}
\isbn{978-1-4503-6819-3/20/04}
\doi{https://doi.org/10.1145/3334480.3383044}

\copyrightinfo{\acmcopyright}

\maketitle

\RaggedRight{} 

\begin{abstract}
Food journaling is an effective method to help people identify their eating patterns and encourage healthy eating habits as it requires self-reflection on eating behaviors. Current tools have predominately focused on tracking food intake, such as carbohydrates, proteins, fats, and calories. Other factors, such as contextual information and momentary thoughts and feelings that are internal to an individual, are also essential to help people reflect upon and change attitudes about eating behaviors. However, current dietary tracking tools rarely support capturing these elements as a way to foster deep reflection. In this work, we present Eat4Thought---a food journaling application that allows users to track their emotional, sensory, and spatio-temporal elements of meals as a means of supporting self-reflection. The application enables vivid documentation of experiences and self-reflection on the past through video recording. We describe our design process and an initial evaluation of the application. We also provide design recommendations for future work on food journaling.
\end{abstract}




\section{Introduction}
In recent years, with the popularity of personal digital devices, an increasing number of self-tracking tools have been created to help people monitor health-related behaviors. Food journaling is one of the commonly used methods that aims to help people identify eating habits and typically requires self-reflection on eating behaviors~\cite{cordeiro2015rethinking, cordeiro2015barriers, doumit2016effects}.

Current solutions mostly focus on tracking nutrient intake~\cite{ferrara2019focused}. These tools typically support a review of food consumption, nutrients, and calories to help users reflect on their eating. Yet, prior work has identified challenges with diet tracking apps that focus on food intake, such as user fatigue in logging~\cite{thomaz2015inferring}, the inaccuracy of nutritional databases~\cite{cordeiro2015barriers}, the difficulty of homemade meal entry~\cite{cordeiro2015barriers}, and little reflection on collected data~\cite{choe2017understanding}. 
Yet, many other factors also influence eating behaviors and how food choices are made. These factors include subjective factors like mood and sensory appeal~\cite{steptoe1995development}, as well as elements that are external to an individual, such as the setting, lighting, temperature, convenience, and who one eats with~\cite{sundbo2008creating}. These contextual features are essential to help people reflect upon and change attitudes about eating behaviors~\cite{grimes2008eatwell}. However, food journaling apps rarely support the capture and reflection upon such information.  

Our work expands upon the traditional food journaling approach, which typically focuses on tracking food intake, to support eating experience documentation that includes broader and richer information collection. In particular, we explore how technology can be designed to help users reflect on contextual information and internal moods and emotions. We designed and developed an eating experience journaling application that supports video recording to capture eating experience through narrating both internal feelings of users and external factors, such as the setting, atmosphere, lighting, and temperature. The app aims to allow users to vividly document emotional, sensory, and spatio-temporal elements of their eating experience to support self-reflection. This work will contribute insights to the design of personal health informatics tools. 

\section{Related Work}

Food journaling is one of the most commonly used approaches to self-monitoring eating behaviors~\cite{burke2008using}. In this section, we begin with a review of food journaling implementation methods: traditional food journaling, wearable devices, automated recording, and photographic journaling.

Traditional food journaling allows users to record food consumption using pencil-and-paper and electronic solutions. Early methods using pencil-and-paper were mostly used in questionnaires and interviewing techniques to record the food intake of a subject~\cite{zepeda2008think}. Paper-based food diaries provide a convenient way of recording diet and help overcome the issues in short-term memory in dietary self-report~\cite{doumit2016effects}. It also encourages people to be aware of eating healthy and to adopt healthier eating behaviors since the food logging process requires self-reflection~\cite{thomaz2015inferring}. More recent techniques using smartphone applications support food consumption entries including nutritional information, such as carbohydrates, calories, fats, and proteins with help of nutritional database~\cite{cordeiro2015rethinking}. 

Prior research has also identified the challenges of using traditional self-monitoring methods. Keeping track of too much information that is not relevant to one's goals is one reason that people feel such dietary tracking is tedious, and thus stopping journaling~\cite{thomaz2015inferring}. Moreover, users often have uncertainty when creating detailed entries, especially for home-cooked dishes~\cite{cordeiro2015barriers}, meals that are prepared by friends, and food ordered in restaurants. It is difficult to calculate the total calories, estimate how much one eats, and track the nutritional information of these meals. Relatedly, the inaccuracy and unreliability of nutritional databases also creates challenges for dietary tracking~\cite{cordeiro2015barriers}. These challenges may further lead to a high user drop-out rate in the first few weeks of using the food journaling tools. 

Wearable devices and automatic sensing have also been used to help people improve dietary behaviors. The Microsoft SenseCam is a wearable device that automatically captures photos of a wearer’s environment at periodic intervals~\cite{chen2013using}. To facilitate reflecting on more contextual information, the SenseCam has been applied in understanding travel behaviors, sedentary behaviors, and estimating nutritional intake~\cite{chen2013using}. However, full automation can undermine the mindfulness and self-reflection needed for eating behavior change~\cite{cordeiro2015rethinking}.  

\begin{marginfigure}[1pc]
  \begin{minipage}{\marginparwidth}
    \includegraphics[width=0.9\marginparwidth]{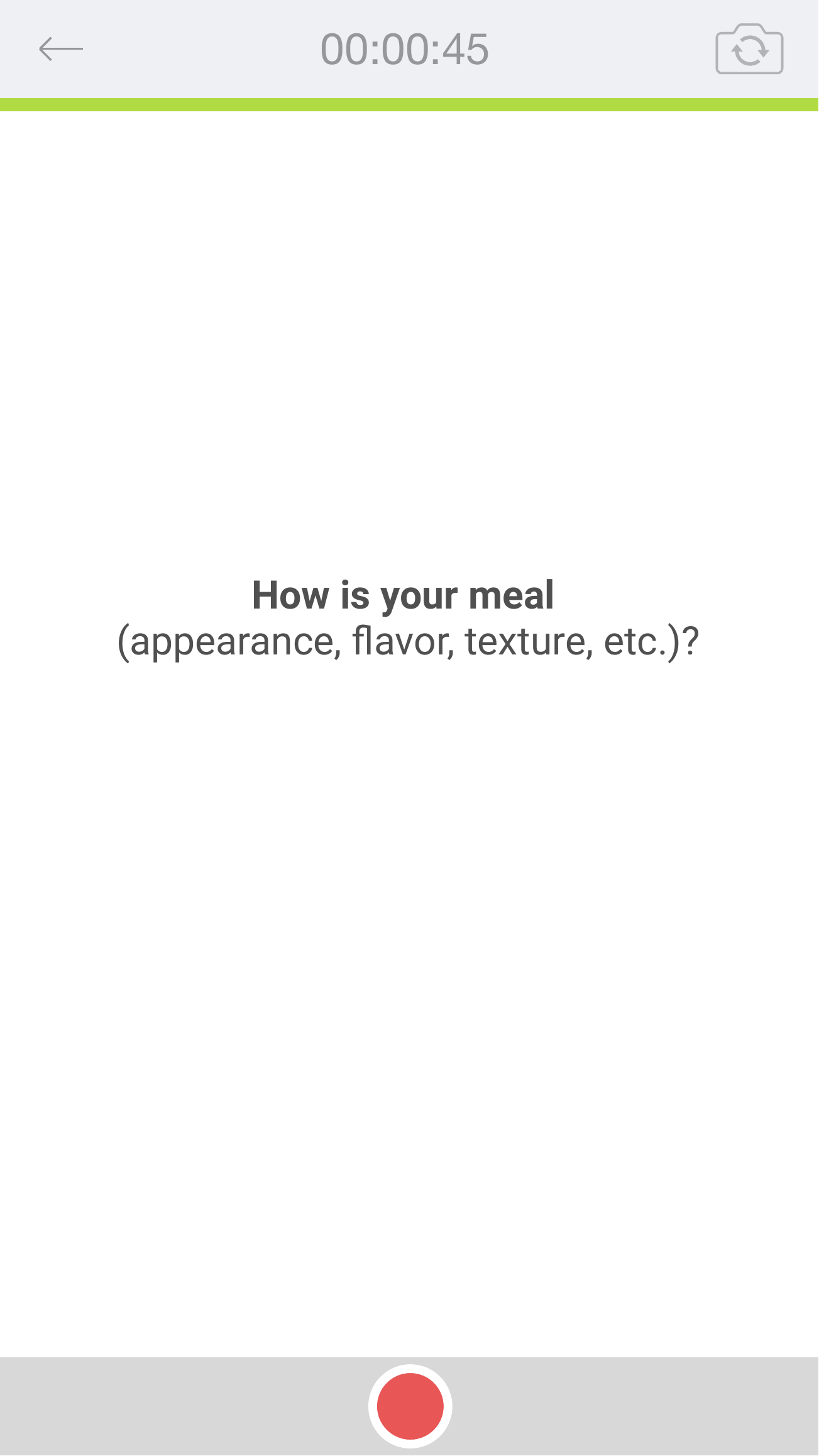}
    \caption{An example question to support reflection on eating experience in the video mode}~\label{fig:video}
  \end{minipage}
\end{marginfigure}

To help address the inaccuracy of food databases and to simplify the food journaling process, prior work has explored the use of photographic food diaries~\cite{higgins2009validation}. Previous qualitative research indicated that photographic food diaries can alter attitudes and behaviors associated with food choices~\cite{zepeda2008think}. Combining food photos and data about the subjective dimensions of experiences and contextual factors (e.g., time, location, and how people feel before eating) helps promote reflection on healthy eating behaviors~\cite{cordeiro2015rethinking}.
Indeed, an increasing amount of work has suggested that self-tracking tools should consider supporting capturing contextual information to facilitate reflection on past experience (e.g.,~\cite{choe2017understanding, cordeiro2015barriers, cordeiro2015rethinking}). 
More recently, audio-video features have been proposed in the design of food journaling tools. Video recording can add more dimensions to the photographic food diaries to further help qualify nutrient consumption~\cite{robertson2017novel}, and enable users to describe any hidden food ingredients while recording videos~\cite{jago2019assessment}. 

However, there has been little work exploring how capturing more contextual information (e.g., factors external and internal to an individual) as well as audio-video methods can be used to help contextualize eating experiences and foster reflection on eating behaviors. Our work aims to address this research gap. 

 
\section{Food Journal Design and Implementation} 

In this work, we present \emph{Eat4Thought}, a food journaling application that allows users to document eating experiences and reflect on contextual factors that are external to an individual (e.g., environmental elements such as setting and temperature), as well as factors that are internal to an individual (e.g., emotions and feelings). 

\subsection{Application Design}

Eat4Thought enables users to take a video of their meal or environment, and narrate to describe how they feel about their eating experience (as shown in \autoref{fig:video}). It also provides three probing questions to facilitate user reflection while recording:

\textit{Q1) How is your meal (e.g., appearance, flavor, texture, etc.)?} \\
Q1 prompts users to describe their meals more vividly regarding their sensory experiences in the mouth and throat, such as flavor and texture.  

\textit{Q2) How is your eating environment (e.g., setting, temperature, music, lighting, service, etc.)?} \\
Q2 emphasizes factors that are external to an individual (e.g., the temperature, music, lighting, and restaurant service), and helps users reflect on how such factors may influence their eating experiences. 

\textit{Q3) How are you feeling about your eating?}  \\
Q3 encourages users to express their internal feelings and emotions. 

The rationale behind the video-aided feature is that video provides a visual-auditory presentation of information and is beneficial to cognitive learning and can foster self-reflection~\cite{cheng2009digital}. The use of video has received more attention in recent years in the areas of education and professional training~\cite{cheng2009digital}. Video provides additional richness beyond that of photographic food dairies, such as capturing eating contexts simultaneously~\cite{jago2019assessment}. We used smartphone video recording to enable rich data capture, and to allow users vividly document their eating experiences and evocatively recall their food choices, external environmental features, and internal feelings. As such, our goal was to help users expressively curate their experience. 

\begin{marginfigure}[1pc]
  \begin{minipage}{\marginparwidth}
    \centering
    \includegraphics[width=0.9\marginparwidth]{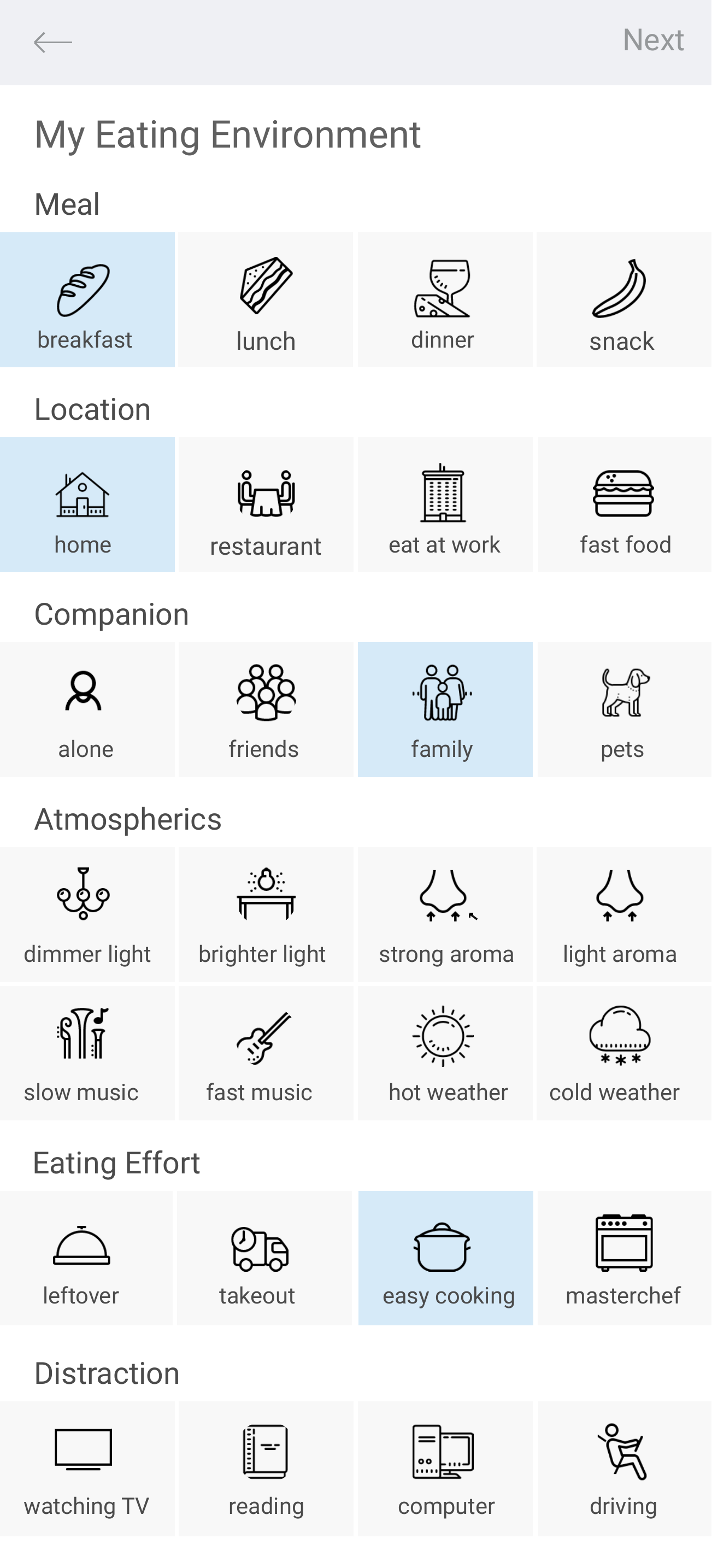}
    \caption{An example of predefined categorized tags about ``My Eating Environment''}~\label{fig:tags}
  \end{minipage}
\end{marginfigure}

After recording the video, Eat4Thought displays a set of predefined categorized tags (e.g., as shown in \autoref{fig:tags}) that are related to the prompted questions (Q1- Q3), including aspects of meal content, eating environment, and emotions and feelings about eating experience. These tags primarily functioned as prompts to help users think differently about their eating behaviors. Users were also encouraged to create customized tags using the app. 

Inspired by the work of Sundbo and Darmer~\cite{sundbo2008creating}, we included the following environment tag categories: meal type, location, companion, atmospherics, eating effort, and distraction (see \autoref{fig:tags}).  Eat4Thought also provides food tags, including types of drink, vegetables, protein, grains, and fruit (see \autoref{fig:food}). Finally, we asked users about their in-the-moment feeling about their eating experience; these tags including positive feelings, such as relaxed and happy, and negative feelings, such as boredom, stressed, and uncomfortable.

\subsection{Implementation}

Eat4Thought was a browser-based web application, implemented with Framework7.io~\cite{framework7}~(a full featured HTML framework for building mobile applications), jQuery, and pouchdb~\cite{pouchdb}. The backend server software was implemented with Node.js. Users can add Eat4Thought to their home screens for a native app experience.  

\section{Method}

\subsection{Evaluation Method}

Upon approval from the Institutional Review Board at our institution, we conducted a qualitative study with eight participants. We posted flyers at our institution and public places around the local town (e.g., at train stations). We met participants individually and had them use the application at a meal. The locations where participants had their lunch or snack include their offices and cafes. After greeting participants and having them sign the consent form, we left them to prepare or purchase their meal. The goal was to minimize the impact of researchers' presence on their food choice. 
 
We first showed the app to participants. We asked participants to verbalize their thoughts as they move through the user interface, using the ``think-aloud'' method~\cite{van1994think}. Specifically, we asked about their initial reactions to and expectations for the app, and explain to us whatever they were looking at, thinking and feeling at each moment, to the best of their ability. We then conducted follow-up interviews with participants, probing topics such as their perspectives on healthy eating and things that may influence their dietary choices and eating experiences. Lastly, we asked participants to complete a survey about their demographic information, and prior experiences with dietary tracking. Each session took approximately 80 minutes. Participants received \$25 USD at the end of the session.

\subsection{Participants}

We recruited eight participants in total for an initial evaluation of our app, including five female and three male participants. The median age of the participants was 30. Five participants were married and two of them had kids. The calculated body mass index (BMI) indicated that all of the participants were in the healthy range with an average BMI 21. Five participants had completed college or university and three held a graduate degree.

\subsection{Data Collection and Analysis}

All sessions were audio-recorded and transcribed. We also video recorded how participants interacted with the app. We conducted an inductive qualitative analysis~\cite{thomas2006general} of the transcripts as well as notes we kept during observations from the study. The first author performed open coding and then iteratively clustered these codes into higher-level themes based on the codes. 
 
\section{Findings and Discussion}

Two themes emerged from our qualitative analysis. Our app was able to help contextualize participants' eating experiences and draw their attention to specific moments from their past experiences to help them reflect on eating behaviors. Eat4Thought also raised some participants’ awareness of healthy eating.

\subsection{Contextualizing Eating Experience} 

\begin{marginfigure}[-20pc]
  \begin{minipage}{\marginparwidth}
    \centering
    \includegraphics[width=0.9\marginparwidth]{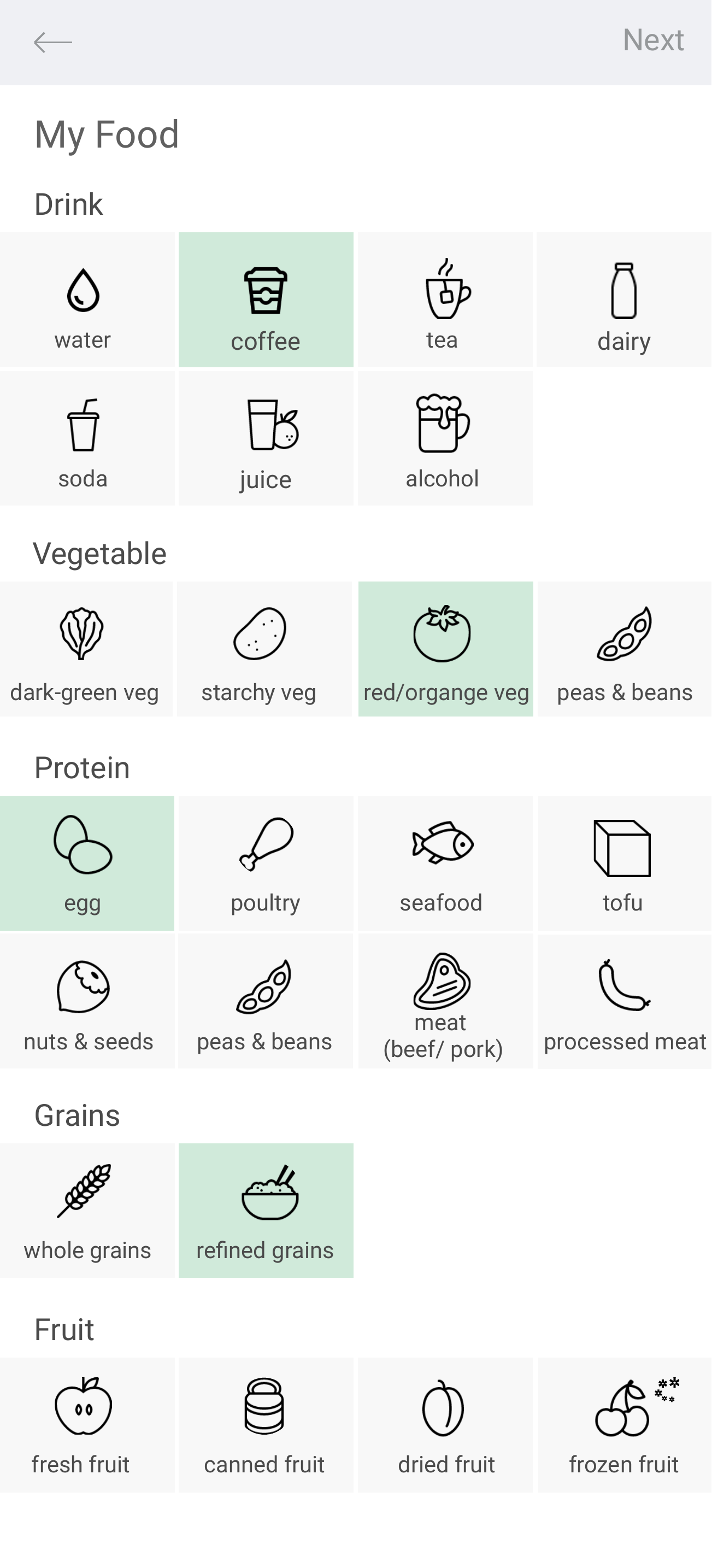}
    \caption{An example of predefined categorized tags about ``My Food''}~\label{fig:food}
  \end{minipage}
\end{marginfigure}

Our participants indicated a few elements that contribute most to memorable eating experiences---people who they eat with and environmental context. 
Participants also realized that how internal factors (e.g., mood) influence one another. Some of them (n=3) talked about how mood and emotions influence their dietary choices. For example, P1 came to realize that she would be more mindful of healthy eating when she was in a good mood. P1 said:
\begin{quote}
    ``\textit{if I’m in a good mood, I’ll choose something I like and healthy, but if I am not happy, maybe I just eat it and I don’t have time to care about my meal}''.
\end{quote}

Eat4Thought supports vivid documentation and reflection on eating experience through visuals and sound using videos. Being able to capture specific moments from their past helped participants view their experiences from different perspectives and scaffold reflection on how previously successful eating experiences contribute to their dietary choices. For example, when P6 was using the video feature to record his meal, he recalled his past eating experience:
\begin{quote}
    ``\textit{That Thai restaurant with very good white curry, very healthy, [and] we liked it. [It was] in Hawaii. I guess our mood was different. You are on vacation, you are in the mood to enjoy the food... But if you are in a very nervous situation, even if you go to a really good restaurant, you may not feel the same}''. 
\end{quote}
P6 later indicated that his past dining experience in Hawaii prompted him to try to cook healthy meals at home using similar ingredients (i.e., white curry). 

\textbf{Implications:}
In general, Eat4Thought explores how food journaling app can be designed to support vivid documentation of eating experience that can go beyond tracking nutrients intakes. Future food journaling tools may consider to support contextualize users' past experiences and help capture specific moments that are important to them in a vivid way, such as using video and audio. Doing so may enable users to vividly curate expressive experience and foster reflection on eating behaviors. 

\subsection{Raise Healthy Eating Awareness}

Our participants indicated that healthy eating knowledge has played a dominant role in determining their dietary choices. Yet, many participants had concerns about a lack of healthy eating knowledge. For example, P7 said she did not have substantial knowledge of healthy eating. 
Our participants indicated that Eat4Thought helped them become more aware of healthy eating. Specifically, tags provided in the ``My Food'' page (as shown in Figure~\ref{fig:food}) include types of food (e.g., vegetable, protein, grains, and fruit).
P5 indicated that she learned about healthy eating from the app \emph{when selecting the food-related tags}.  
\begin{quote}
    \textit{``I just learned from this app... I saw different categories... Hmmm, my meal only contains grains basically... I think I should try all these categories''}.
\end{quote}
Similarly, when selecting the categories of their food, P5 asked herself:
\begin{quote}
    \textit{``healthy protein? Does it mean if you eat some protein but is unhealthy? I think like red meat?}''.
\end{quote}
Prompted by the tags of food categories, Eat4Thought helped our participants raise questions about their food consumption. Many other participants like P5 expressed some level of uncertainty regarding the healthiness of their food and how to categorize their food using the tagging system provided by the app. 

Moreover, selecting contextual tags helped our participants reflect on how interpersonal relationships impact their eating behaviors. Particularly, participants who were married and had kids indicated that their dietary choices for family meals primarily focused on the nutritional needs of their kids. For example, when selecting the tags (i.e., ``companion''), P1 came to realize she would first make sure to cook a healthy meal for her baby, as a new mom. But she did not pay attention to if she and her husband eat healthfully. 
We also found that the keyword ``mom'' was mentioned often when talking about how their healthy eating knowledge comes from. Rarely, participants mentioned ``dad'' or the role of ``male'' in passing the knowledge down to the young generation. Indeed, the definitions and awareness of healthy eating vary in context among people. Individual interpretation of a healthy diet is influenced by knowledge, environmental factors, informational sources~\cite{falk2001managing}, beliefs and values, self-efficacy, and socio-economic status (SES)~\cite{inglis2005women}. 

\textbf{Implications:}
Our findings suggest that contextual tagging systems can help promote reflection on eating behaviors. An interesting extension of this work could be to examine what contextual tags best motivate reflection in order to design tagging features more effectively.  
More research is needed to further explore gender differences and family dynamics (e.g., caregiver roles) when designing and evaluating self-tracking tools. Moreover, we suggest future research to further explore 
1) how to help users identify and create personalized healthy-eating contextual tags to promote reflection and reconstruct experiences, and 
2) how contextual prompts can be designed considering people may have varied priorities in managing family health (e.g., a focus on kids' needs to increase motivation).

\section{Limitations and Future Work}

Our evaluation study included eight participants at one meal due to the time constraint. Recruiting more participants would help generalize our findings. 
Future work should also evaluate how well video-aided storytelling methods can encourage long-term user engagement and healthy eating.

\section{Conclusion}

In this work, we contributed a food journaling application, Eat4Thought, which aims to help people vividly document their eating experience that includes both contextual information and momentary thoughts and feelings that are internal to an individual. Through this initial qualitative study, we found that the app helped participants to reconstruct past eating experiences, through vivid documentation of these experiences using video and tagging systems. Our app also helped users raise awareness of the importance of healthy eating. 

\section{Acknowledgements}
We thank all participants for their participation in our study. We also thank our anonymous reviewers for their feedback.
 
\balance{} 

\bibliographystyle{SIGCHI-Reference-Format}
\bibliography{sample}

\end{document}